# LIPSS pattern induced by polymer surface instability for myoblast cell guidance

**Nikola Slepičková Kasálková[a], Veronika Juřicová[a], Silvie Rimpelová[c], Dominik Fajstavr[a], Bára Frýdlová[a], Zdeňka Kolská[b], Václav Švorčík[a], Petr Slepička[a,*]**

[a]Department of Solid State Engineering, The University of Chemistry and Technology Prague, 166 28 Prague, The Czech Republic

[b]Faculty of Science, J. E. Purkyně University in Ústí nad Labem, 400 96 Ústí nad Labem, The Czech Republic

[c]Department of Biochemistry and Microbiology, The University of Chemistry and Technology Prague, 166 28 Prague, The Czech Republic

**Abstract**

The presented study highlights the efficiency of employing a KrF excimer laser to create diverse types of periodic nanostructures (LIPSS – laser induced periodic surface structures) on polyether ether ketone (PEEK) and polyethylene naphthalate (PEN) substrates. LIPSS structures are very important both in tissue engineering and find also strong application in the field of sensor construction, and SERS analysis. By exposing the polymer films below their ablation threshold to laser fluence ranging from 4 to 16 mJ·cm$^{-2}$ at 6,000 pulses, we studied both single-phase exposure at beam incidence angles of 0° and 45°, and two-phase exposure. Atomic force microscopy analysis revealed that the laser-treated samples contained distinctive periodic patterns such as waves, globules, and pod-like structures each exhibiting unique surface roughness. Moreover, using analytical methods like EDS and XPS shed light on the changes in the atomic composition, specifically focusing on the C and O elements, as a result of laser exposure. Notably, in almost all cases, we observed an increase in oxygen percentage on the sample surfaces. This increase not only led to a decrease in the contact angle with water but also lowered the zeta potential value, thus showing that the modified samples have enhanced hydrophilicity of the surface and altered electrostatic properties. Last but not least, the samples were assessed for biocompatibility; we studied the interaction of the prepared replicates with mouse myoblasts (C2C12). The impact of globular/dot structures on the cell growth in comparison to pristine or linear LIPSS-patterned surfaces was determined. The linear pattern (LIPSS) induced the myoblast cell alignment along the pattern direction, while dot/globular pattern even enhanced the cytocompatibility compared to LIPSS samples. Through this comprehensive analysis, the research underscores the multifaceted implications of employing KrF excimer laser-induced nanostructures, ranging from surface morphology alterations to biocompatibility enhancements, thus, opening new avenues for advanced material engineering.

*) Corresponding author:

P. Slepička, tel.: +420 220 445 162; E-mail address: petr.slepicka@vscht.cz

## 1. Introduction

Although hundreds of polymers can be easily synthesized and used as biomaterials, only 10–20 polymers are used to make disposable medical devices and long-term implants [1]. An overview of selected polymeric biomaterials, including their applications involve, for instance, polymethyl methacrylate (PMMA), poly(2-hydroxyethyl methacrylate) [PHEMA], polyethylene terephthalate (PET), polytetrafluoroethylene (PTFE), polyurethanes (PU), silicones, and other polymers [2]. However, in tissue engineering and regenerative medicine, the main attention has been focused on natural biopolymers and synthetic biodegradable polymers [3]. Natural biopolymers such as polypeptides, proteins, and polysaccharides usually have a very high degree of biocompatibility. As for proteins, the most used ones for biopolymer preparation are collagen, gelatin, fibrinogen, elastin, keratin, actin, and myosin, while polysaccharide-based polymers are mostly composed of cellulose, dextran, chitosan, alginate, chitin, and glycosaminoglycans [4]. Nevertheless, their advantageous properties, such as biodegradability can be counteracted with uneasy chemical modifiability, and also, some modifications may result in changes in important properties of the biopolymers [5]. Conveniently, recent advances in material synthesis and processing technologies have opened up new avenues for improving biocompatibility and other material properties that were previously unattainable [6]. Even the important role in tissue engineering application play biopolymers, also aromatic polymers may be used either for cytocompatibility improvement or cell guidance.

One of the most important representatives of aromatic polymers, including polyaryletherketones (PAEK), as semi-crystalline thermoplastics, is polyether ether ketone (PEEK). It is composed of very stable aromatic structural units which give this polymer high chemical resistance, including resistance in organic and aqueous environments, except for 98% sulfuric acid [7]. In addition to the PEEK chemical structure, its behavior is affected by its supramolecular arrangement, mainly by the ratio of the crystalline and amorphous phases [8]. The content of the crystalline phase varies depending on the heat treatment, with the melting temperature ($T_m$) of ca.

343 °C at which most of the crystalline phase dissolves [9]. The research reported on PEEK and its composites has shown that it has long-term stability in an aqueous environment at 37 °C and, thus, it became a suitable candidate for biomaterial applications, especially in the field of bone tissue engineering [10]. In addition, PEEK utilization is convenient also for its high resistance to the degrading action of ionizing radiation and, compared to metal and ceramic biomaterials, it has a more suitable value of Young's modulus of elasticity. The elastic moduli of reinforced composites, such as CFR-PEEK (carbon fiber reinforced PEEK), can be as high as 18 GPa [11], bringing them significantly closer to the value of cortical bone (19.9 ± 1.8 GPa) [12]. Due to the relative inertness of PEEK-based biomaterials, among other things, they are also an attractive platform for novel bioactive material development. In this direction, some studies have already been done with the addition of hydroxyapatite (HA) and tricalcium phosphate (TCP) [13, 14]. Currently, however, PEEK has the greatest clinical impact in applications of spinal implant construction and orthopedic replacements. Further, at an advanced stage of clinical testing, there are also applications in dental medicine [15].

The next important representative of aromatic polymer is polyethylene naphthalate (PEN). Unlike PET, polyethylene naphthalate has a higher glass transition temperature ($T_g$) of approximately 112–120 °C [16]. By adding PEN to PET by creating a PET-PEN composite, the $T_g$ of the composite increases, which is desirable when sterilizing bottles and making jams. PEN also has a higher barrier property for oxygen and carbon dioxide, thus, making it more suitable for food and beverage packaging [17]. The high PEN melting point ($T_m$) ranging between 250-265 °C enables its use as a structural material for various industrial sectors. Recently, so-called ferroelectrics have also been successfully developed from PEN films, which were previously produced from polypropylene, PET, or cycloolefin copolymers [18]. Advantageously, PEN is also biocompatible, has a hydrophilic surface, is optically transparent, and is UV-resistant. Specific cases of biomedical applications include stents, drug delivery carriers, biological sensors, and microfluidic devices [19-22]. One of the other promising applications may also be the use of PEN substrates for the production of biocompatible sensors for the detection of biomarkers of inflammatory diseases, which can detect several biomarkers simultaneously [23, 24].

Laser micro- and nanostructuring of materials represent a significant contribution in many fields of science, technology, and medicine, such as the production of optical and nanoelectronic devices [25], optical memory systems [26], and sensors for determining the physical and optical properties of solids in biomedical engineering [27]. The development of technologies for the synthesis of microstructures and nanostructures has been supported by recent advances in laser engineering, which have made it possible to generate pulses with extremely short durations, in the order of femtoseconds (10-15) and picoseconds (10-12). The most frequently used ones are excimer lasers with a wavelength of 193 and 248 nm (ArF and KrF, respectively) [28, 29]. In recent years, there has also been a growing interest in diode lasers generating nanosecond pulses, which offer the advantages of increased beam quality, high efficiency, and reduced space requirement of the device [30-32]. Laser nanostructuring represents a suitable alternative for the creation of high-resolution structures not only on polymer materials but also on metal and glass materials. Examples are laser ablation-based techniques (e.g. laser foaming (LF) [33], laser beam lithography (LIL) [34], laser writing technology (LIFT) [35], pulsed laser deposition (PLD) [36], assisted matrix pulsed laser evaporation (MAPLE) [37], and laser-induced periodic surface patterning (LIPSS) [38]. Nanostructures created by laser modification of materials often have unique properties that cannot be easily achieved by other methods [39]. Several types of LIPSS pattern has been prepared on aromatic polymers, ranging from a periodic linear pattern to dots or even more complicated structures [40-50], its biocompatibility have been reported for PET or PS e.g. in [51,52], the antibacterial studies were also conducted [54]. Also "two phase modification" have been reported successful of several solid substrates [55-58]. Such surfaces created by laser nanostructuring have many interesting properties, for instance, improved adhesion to biological structures, increased ability to absorb biomolecules, or improved tribological properties. One of the techniques for effective surface modification of polymers is their irradiation with an excimer laser. The term "excimer laser" does not describe a single device, but a group of pulsed gas lasers that emit pulses of ultraviolet light radiation in the 157-351 nm region [59].

In this paper we have aimed on the exposure of typical aromatic polymers which find applications in tissue engineering, PEN and PEEK polymers. The surface chemistry and morphology was significantly altered by the excimer laser exposure, which led in a certain interval in construction of periodic pattern so called LIPSS. The novelty of this approach is based on the multiple exposure, which led to formation of specific superposed patterns, such as globular pattern superposed on LIPSS. The main aim was both to increase the polymer's cytocompatibility, and also guide the growth of myoblast cells along the prepared linear periodic morphology, where both of our aims were fulfilled.

## 2. Experimental

*2.1 Materials used*

For laser nanostructuring, PEEK (amorphous) and PEN films with a thickness of 50 μm and ρ of 1.26 and 1.36 g·cm$^{-3}$, respectively, were used. Their highest working temperatures are 250 and 116 °C, respectively. Both polymers were supplied by Goodfellow Cambridge Ltd., UK.

*2.2 Laser exposure*

A KrF excimer laser with a Nova Tube tube with a wavelength of 248 nm and a pulse duration of 20 ns (COMPexPro 100K, Coherent, Inc., USA) was used to irradiate the surface of two biocompatible polymers, PEEK and PEN. The laser beam was linearly polarized by a fused quartz cube that served as a polarizing filter for UV radiation. The cube had dimensions of $25\times25\times25$ mm$^3$ and was equipped with an active polarization layer made of $MgF_2$. The laser irradiation of the substrates was carried out under different conditions, in which parameters such as the angle of incidence and the energy of the laser beam varied to create periodic ripples on the nanometer scale. Single- and two-phase excimer exposures were also observed, whereby the sample was rotated at 90° in the two-phase exposure to create a so-called cross-modification.

To achieve homogeneous irradiation of the samples, only the middle part of the laser beam was used, by placing an aperture with a hole of 0.5 x 1.0 cm² or 1.0 x 1.0 cm². The optical path of the beam was adapted with $SiO_2$ dielectric mirrors for a combined 248/633 nm reflection. The samples were mounted on a Performance 07OTM501 sliding optical stage. To create nanostructures, PEEK and PEN foil samples were used, which were irradiated with a laser at incidence angles of 0° and 45°. The modification took place at laser fluxes in the range of 4 to 16 mJ·cm$^{-2}$ with the help of an attenuator. The laser exposure time corresponded to 6,000 pulses and a frequency of 10 Hz at room temperature (RT, 24 °C). The number of pulses in each of laser phases before and after rotating the sample is the same (the same laser energy dose before and after the rotation of the sample). We will use the term cross-modification / two phase modification further in the paper, the meaning is two processing steps with crossed polarization directions.

*2.4 Analytical methods*

Surface morphology and roughness of the materials were examined by atomic force microscopy (AFM) using Dimension ICON (Bruker Corp., Billerica, MA, USA). The samples were analyzed in Scan-Assyst® mode using nitride lever SCANASYST-AIR with Si tip (spring constant of 0.4 N·m$^{-1}$). NanoScope Analysis software was applied for data processing. Surface roughness ($R_a$) represents the arithmetic mean of the absolute values of the height deviations measured from the central plane.

The morphology of the sample surfaces was also characterized by a complementary technique using the scanning electron microscope FIB-SEM LYRA3 GMU (Tescan. Brno, Czech Republic). The acceleration voltage was set to 10 kV. To ensure the conductivity of the samples, their metallization was performed using the sputtering technique (Quorum Q300T) by deposition of a Pt layer (the thickness of 10 nm, Pt target, purity of 99.9995%). The elemental composition was measured by energy-dispersive X-ray spectroscopy (EDS, analyzer X-ManN, 20 mm$^2$ SDD detector, Oxford Instruments, United Kingdom), while the accelerating voltage for SEM-EDS analysis was set to 10 kV.

The elemental composition on the material surface was analyzed by X-ray photoelectron spectroscopy (XPS) using a spectrometer ESCAProbeP (Omicron Nanotechnology Ldt., Taunusstein, Germany). As a source, a monochromatic X-ray at an energy of 1486.7 eV was used. Atomic concentrations of elements were determined from the individual peak areas using CasaXPS software.

The wettability of the studied samples was determined by measurement of contact angles (CA, θ) on goniometer Advex Instruments (Brno, Czech Republic) connected to the SEE System 7.1 program. Analysis of CA was performed at room temperature with 8 μL drops of distilled water (dyed with methyl violet) using a Transferpette® automatic pipette (Brand, Wertheim, Germany) at 6 different positions of 3 samples in parallel and perpendicular directions. Subsequently, the drops were photographed and evaluated by 3 marked points.

Fourier transform infrared spectroscopy (FTIR) analysis was performed with a Thermo Scientific Nicolet iS5 spectroscope for analysis. The measurement took place in the range of 4,000–525 cm$^{-1}$. Visual representation of graphs was processed using OriginPro 8.5 software.

The zeta potential measurement was carried out using the SurPASS Instrument, using the Helmholtz-Smulochowski method, which enables measuring the zeta potential of solid surfaces. The samples were placed in an adjustable cell where they were in contact with a KCl electrolyte with a concentration of 0.001 mol·dm$^{-3}$. All samples were mounted in two 20×10 mm$^2$ holders and measured at the same pH (∼6).

*2.5 Cytocompatibility testing*

To evaluate the PEEK and PEN sample's cytocompatibility, mouse myoblasts C2C12 were chosen. The cells were regularly passaged at the exponential phase of growth, the detachment was performed by trypsin-EDTA solution. As C2C12 cell cultivation media, Dulbecco's Modified Eagle's Medium (DMEM; Thermo Fisher Scientific, USA) supplemented with 10 % (*v/v*) fetal bovine serum (Thermo Fisher Scientific, USA) was used. The cells were maintained in a cell culture incubator at 37 °C, 5 % $CO_2$ in the atmosphere, and 95 % humidity.

For cytocompatibility measurements, the samples were first immersed in 70 % (*v/v*) ethanol in water (1 h) for sterilization. Then, they were rinsed with sterile phosphate-buffered saline (PBS, pH of 7.4), after which, the tested materials were placed in 12-well plates (VWR, USA) and weighted by PMMA cylinders (Zenit, The Czech Republic). This was followed by C2C12 cell seeding (1.5·10$^4$ cells per 1 cm$^2$) in 1 mL of cell culture media containing. Pristine PEEK and PEN as well as tissue-culture polystyrene were used as controls. All samples were prepared in triplicates.

After 1, 3, and 6 days of incubation, the C2C12 cells were (after double rinsing with PBS) fixed with a 4 % (*v/v*) formaldehyde/PBS solution for 20 min. at 24 °C. After that, the fixative was discarded, the cells were washed with PBS twice and fluorescently labeled with DAPI (Merck, USA; 1 µg·mL$^{-1}$) and conjugate of phalloidin with Atto488 (AttoTec GmbH, Germany; 3 µg·mL$^{-1}$) for 15 min. at 24 °C in the dark. Then, the staining solution was removed, and samples were washed with PBS twice and microscoped.

To examine the number and morphology of C2C12 cells adhered and growing on PEEK and PEN samples, fluorescence microscopy (Olympus IX-81, Olympus, Japan; xCellence software) was used. The cells were imaged with a 10× dry objective (NA of 0.30; total magnification of 100×) using a triple quad filter DAPI/FITC/TRITC (Olympus, Japan) and 150-W xenon arc burner. From each sample, 10 regions of interest were captured by an EM-CCD camera (C9100-02, Hamamatsu, Japan). The images were then background-corrected and DAPI and FITC fluorescence channels were merged.

The cell number was assessed by their nuclei counting with ImageJ software as follows: image binarization, thresholding, applying watershed algorithm and using "analyze particles" function. From the gained data, averages and SD were calculated..

### 3. Results and discussion

*3.1 Surface morphology*

The polymer exposure with different laser fluencies had a significant effect on the formation of nanostructured surfaces. In general, it can be stated that higher laser beam energies were needed to create LIPSS (Figure 1). This fact is not surprising and has been verified in our previous research [41,42], however, distinct laser equipment has been used. The maximum energy of the laser beam, when LIPSS formation still occurred, was found at the energies of 12–14 mJ·cm$^{-2}$. At energies higher than 16 mJ·cm$^{-2}$, gradual distortion of periodic structures already occurred. The values of dimension of ripple pattern are introduced in Table I.

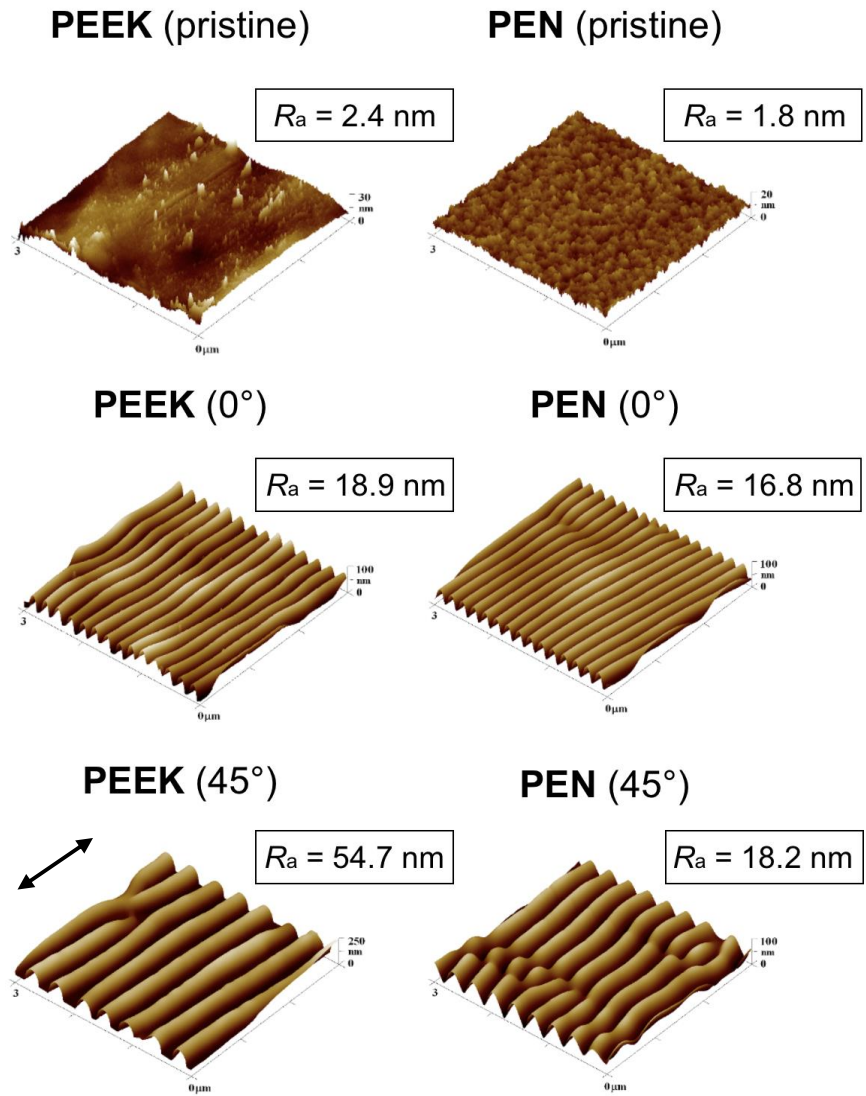

**Figure 1** 3D AFM images (3 × 3 μm$^2$) of both polymer (PEEK, PEN) samples unmodified (pristine) and modified (at an angle of 0° and 45°) with a laser fluence of 10 mJ·cm$^{-2}$ with the number of pulses 6000. To the individual samples the respective values of the mean roughness R$_a$ of their surfaces are given. The arrow indicates the laser polarization direction.

**Table 1** Average surface roughness values ($R_a$) of PEEK and PEN polymer films exposed to a laser fluence of 10 mJ·cm$^{-2}$ (6000 p.), their dependence on the average periodicity value ($\Lambda$) and the average height (h) of the structures. Size of the measured area $3 \times 3$ μm$^2$.

| Sample | $R_a$ [nm] | $\Lambda$ [nm] | h [nm] |
| --- | --- | --- | --- |
| PEEK (pristine) | 2.4 | - | - |
| PEEK (0°) | 18.9 | 218.0 ± 8.0 | 52.7 ± 4.3 |
| PEEK (45°) | 54.7 | 416.8 ± 11.7 | 136.9 ± 6.0 |
| PEN (pristine) | 1.8 | - | - |
| PEN (0°) | 16.8 | 201.2 ± 4.5 | 52.9 ± 3.7 |
| PEN (45°) | 18.2 | 340.6 ± 16.1 | 67.9 ± 2.5 |

When focusing only on the results of excimer laser exposure at the angle of 0°, we find that it had a similar effect on the morphology changes on both substrates (PEEK and PEN), while the ripple formation on these materials was detected only at the laser flux of 6 mJ·cm$^{-2}$. A theoretical explanation of how significantly the surface of this polymer is affected can be found in the absorption coefficient of the given material. In our case, PEN was used with a higher absorption coefficient of 20 μm$^{-1}$ (at 248 nm) and a lower ablation threshold of 40 mJ·cm$^{-2}$ [60] than the absorption coefficient of PEEK, which is ca. 10 μm$^{-1}$ (at 248 nm) with an ablation threshold of 50 mJ·cm$^{-2}$ [61].

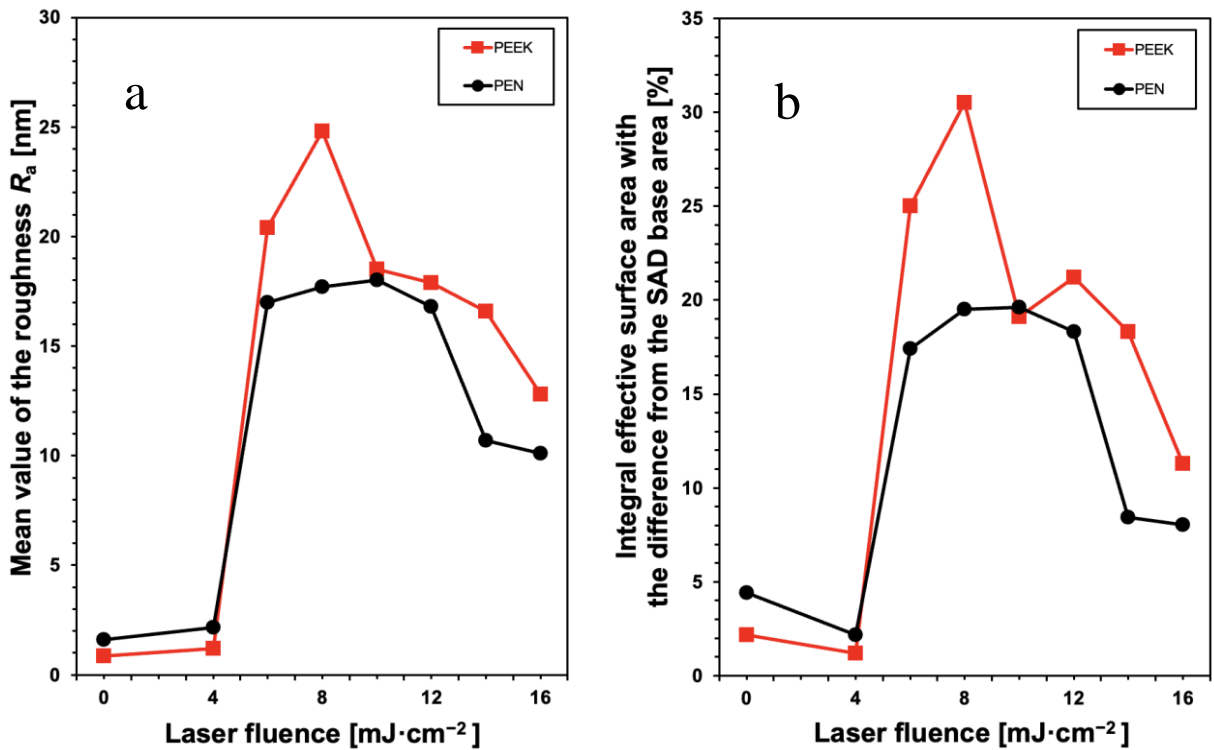

**Figure 2** (a) The dependence of the mean value of the roughness $R_a$ on different fluences of the laser fluence at an angle of 0°. (b) Graph of the dependence of the integral effective surface area with the difference from the SAD base area in units of % on the laser fluence (0°). Measured area $1 \times 1$ μm$^2$. The applied number of laser pulses was 6000.

Figure 2 confirms the trend that the highest difference from the basic surface area (SAD), as well as the surface roughness ($R_a$), are shown by the significantly wavy structured samples prepared at laser fluence ranging from 8 to 12 mJ·cm$^{-2}$. In this context, the behavior of PEEK foils came closest to PEN preparations at a laser fluence of 10 mJ·cm$^{-2}$, when their values were almost identical. The PEEK sample, which was exposed to a laser fluence of 8 mJ·cm$^{-2}$, showed the highest roughness of all monitored substrates. Higher laser fluences have already led to a reduction in roughness and, thus, also to a reduction in the success rate of creating LIPSS structures. In the case of laser exposure at the angle of 45°, the occurrence of LIPSS nanostructures was recorded only at higher values of laser fluences above 7 mJ·cm$^{-2}$. According to the results of SEM and AFM (Figure 3), the PEN substrate was more suitable for the propagation of waves, with its average surface roughness values close to the values observed when exposed at the angle of 0° (~15.9 nm for the scanned area of 3×3 μm$^2$). In contrast, PEEK showed significantly higher roughness values (>43.3 nm for a 3×3 μm$^2$ scanning area).

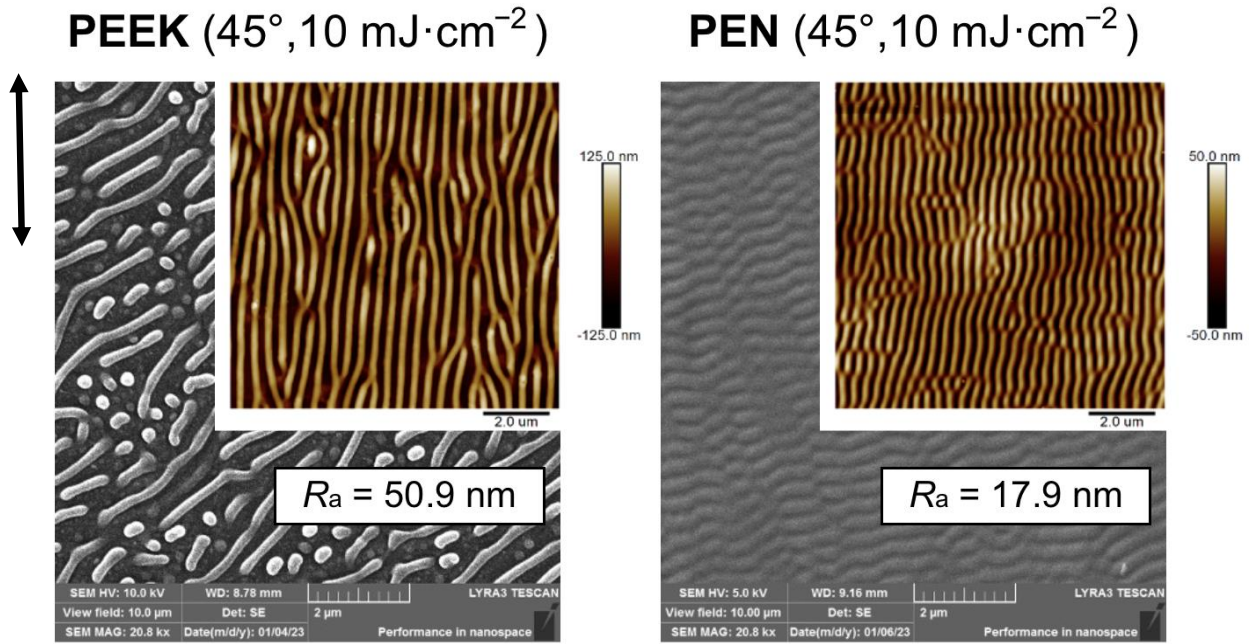

**Figure 3** SEM and AFM images with respective mean surface roughness values of PEEK, PEN polymers. The size of the scanned area is 10 × 10 μm$^2$. The arrow indicates the laser polarization direction. The applied number of laser pulses was 6000.

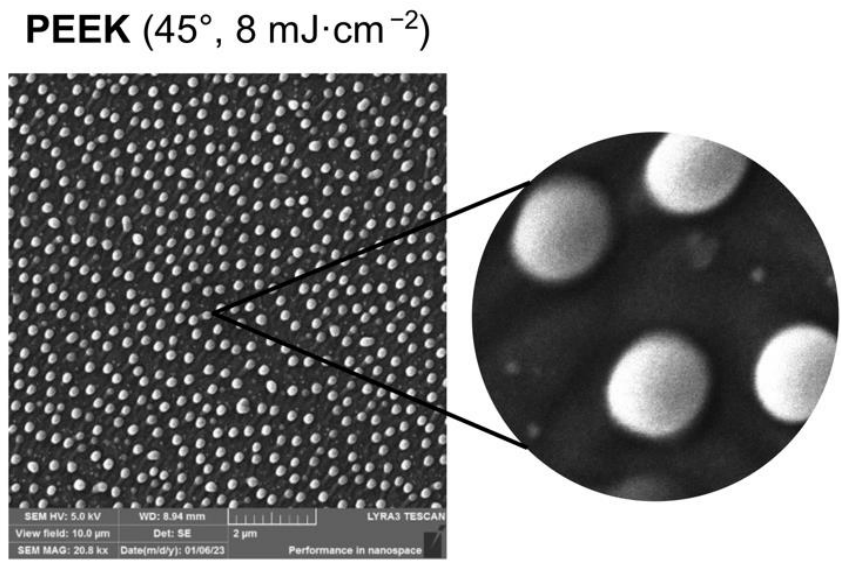

**Figure 4** SEM image (10 × 10 μm$^2$) of the dotted PEEK surface, the laser exposure with 8 mJ cm$^{-2}$ at angle of incidence 45° was performed. The applied number of laser pulses was 6000.

On the PEEK substrate after laser modification at the angle of 45°, areas resembling dots were observed (Figure 4). The origin of the dotted pattern was described by Sendova *et al.*, in his study, the pattern was presented as the result of the superposition of two separate wavelet structures. Similar results were obtained with XU-293 polyimide films [62]. Another observation was focused on two-phase laser modification. As in single-phase laser modification, the morphological behavior of polymer surfaces in cross-modification is similar. However, the resulting appearance of the LIPSS structures was different, and the nanostructures with an exceptional pattern resembling a pod-like structure were formed on PEN (Figure 5).

The first irradiation of the sample induces the linearly formed periodic LIPSS. The second exposure, with perpendicular rotation of the sample induces the formation of superposed/combined globular structures in between the linear structure. We have also analyzed the surface parameters of granular structures, which were formed on the surface during cross modification. The globular structures on PEN surface, which formed due to secondary exposure, has in general smaller geometrical parameters compared to the LIPSS pattern. Its height slightly varied from 20 to 30 nm (for laser fluences from 6 to 14 mJ.cm$^{-2}$), while the diameter of the globular structure was in range between 100-120 nm. The relatively smaller period is also slightly affected by the fact that for LIPSS period the "bottom-bottom" distance of linear pattern was determined, however the overall geometrical parameters are smaller, for PEEK the comparison with similar result has been performed.

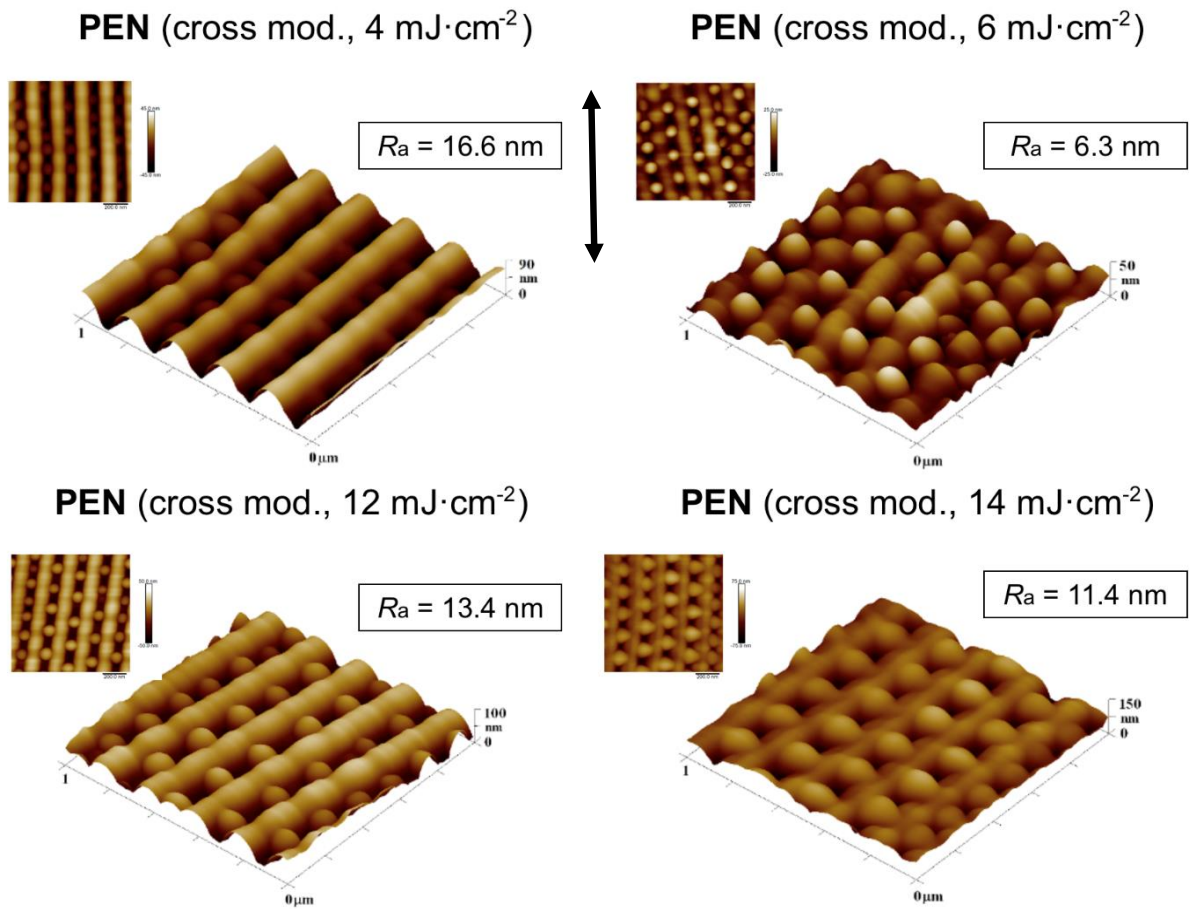

**Figure 5** AFM images of nanostructures on PEN substrates after cross laser modification (4, 6, 12 and 14 mJ·cm$^{-2}$) resembling a pod shape. Scanned area 1 × 1 μm$^2$. The arrow indicates the laser polarization direction. The applied number of laser pulses was 6 000.

Next, the AFM images show that the two-phase modification of the surfaces of the PEN substrate causes the newly formed globular formations (or spheres) to fit into the original wave structure. The most regular distribution of the formed pod-like patterns was observed at laser fluence of 12 and 14 mJ·cm$^{-2}$, lower laser fluences showed a higher diversity of sphere sizes, while at the lowest fluence of 4 mJ·cm$^{-2}$, no spheres were visible. Such multi-level structures containing both larger and smaller structures are also in great demand for tissue engineering purposes, as they support cell adhesion to the substrate. Changes in surface roughness were plotted in Figure 6. The observed trend is different from the previously observed trend for single-phase exposure (see Figure 2). Increasing the laser fluence led to an increase in the value of $R_a$ for both monitored substrates, while the decrease no longer occurred. Nevertheless, there were also variations in roughness between individual samples of the same laser fluence value, with certain areas on the laser-treated PEEK surfaces showing a globular structure instead of a wavy structure (Figure 7).

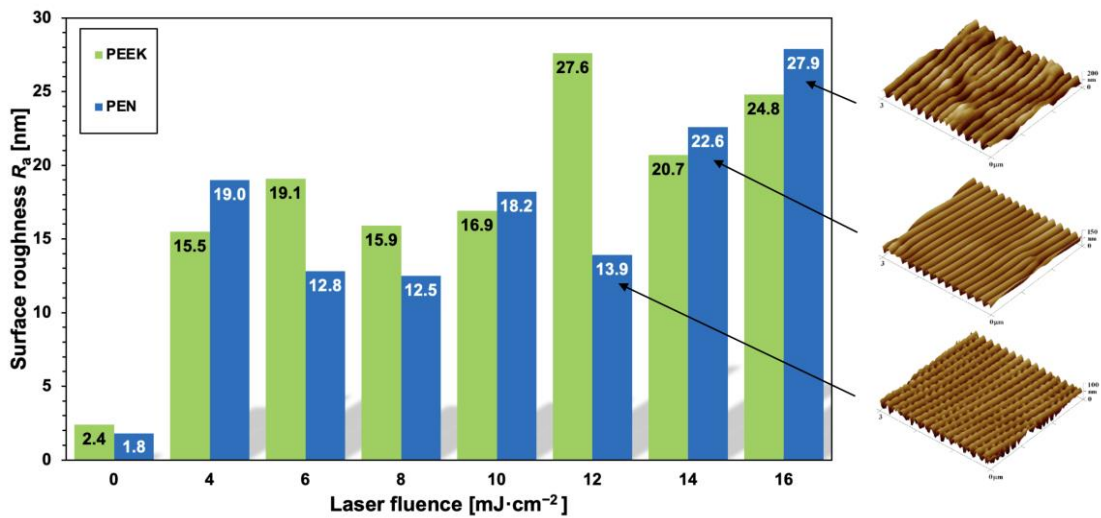

**Figure 6** Dependence of surface roughness of two-phase modified surfaces of PEEK and PEN substrates on laser fluence (6000 pulses). Size of the scanned area 3 × 3 μm$^2$ with relevant images of PEN samples.

*Slepičková Kasálková et al., LIPSS pattern induced by polymer surface instability for myoblast cell guidance*

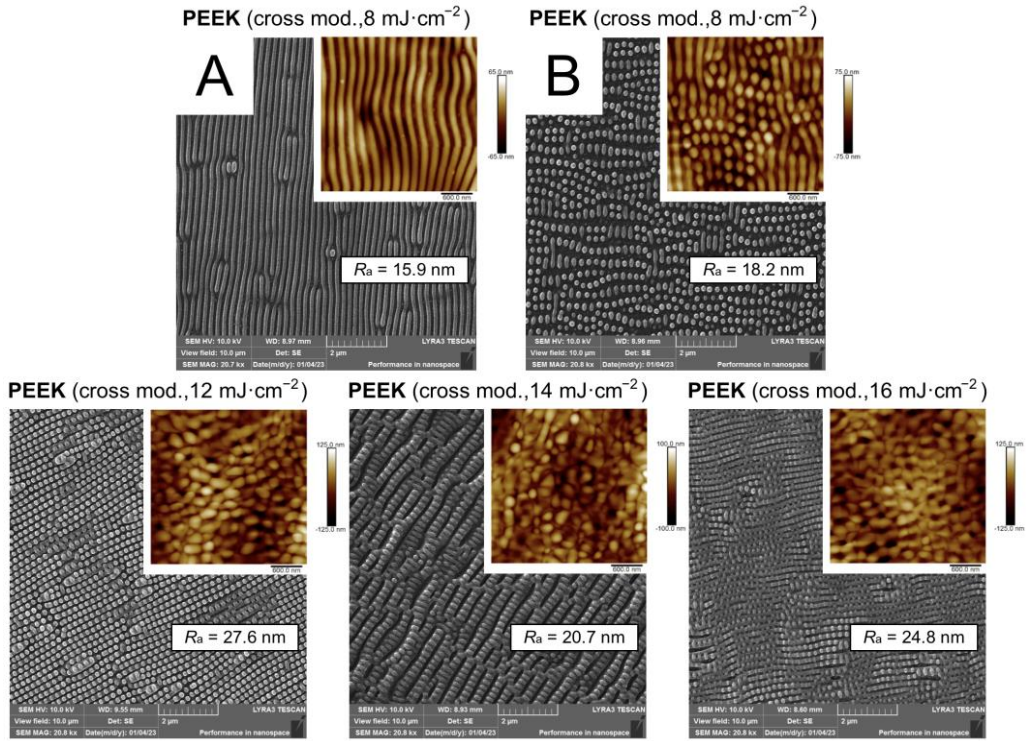

**Figure 7** Dependence (8, 12, 14 and 16 mJ·cm$^{-2}$). Areas A and B of the sample irradiated with a flux of 8 mJ·cm$^{-2}$ serve to compare the appearance of the wavy and globular shape. Image area size SEM 10 × 10 μm$^2$, AFM 3 × 3 μm$^2$. The applied number of laser pulses was 6000.

*3.2 Surface chemistry*

To assess the surface chemical structure of PEEK and PEN films, the elemental composition of the surfaces, their wettability, and electrokinetic potential were studied. The elemental composition of the laser-modified samples was studied using EDS and XPS analysis. By comparing all three types of laser modifications (0°, 45° and cross-modification), based on the initial results of EDS measurements (Table 2), it was found that the highest values of oxygen atomic representation were mainly shown by samples exposed at a laser beam angle of 45°. This trend was observed for both PEEK and PEN substrates (for perpendicularly treated PEEK samples see Figure 8). However, a different trend was observed with two-phase excimer exposure, where the opposite trend appeared. While the cross-modification on the PEEK samples led to the smallest standard deviation differences (σ = ± 0.8 at.%) between the individual laser fluences (4–16 mJ·cm$^{-2}$), the highest fluctuations occurred on the PEN samples (σ = ± 2.6 at.%) across all types of laser exposure. This trend was also observed for PEEK using XPS analysis.

The laser exposure induces inhomogeneous distribution of mass over the treated surface, while the increase of oxygen concentration was observed, in general. But, for PEN we see large increase of oxygen concentration even for lower laser fluencies, so even for 4 mJ.cm$^{-2}$ the increase from 9.5 at. % to 17.1 at. %. A similar increase of oxygen concentration was observed for treatment under angle 45°. The cross modification increase for PEN the oxygen concentration to 13.6 at. %. Further increase of laser fluence increases the oxygen concentration, but due to the mass distribution over the surface, the primary phenomenon of nanopattern formation, the increase is much gradual. The reason is that the oxygen containing groups induced on the surface are re-arranged into the polymer bulk. For PEEK we have also determined the increase of oxygen concentration due to the excimer exposure, for perpendicular exposure, exposure under the angle 45° and cross modification.

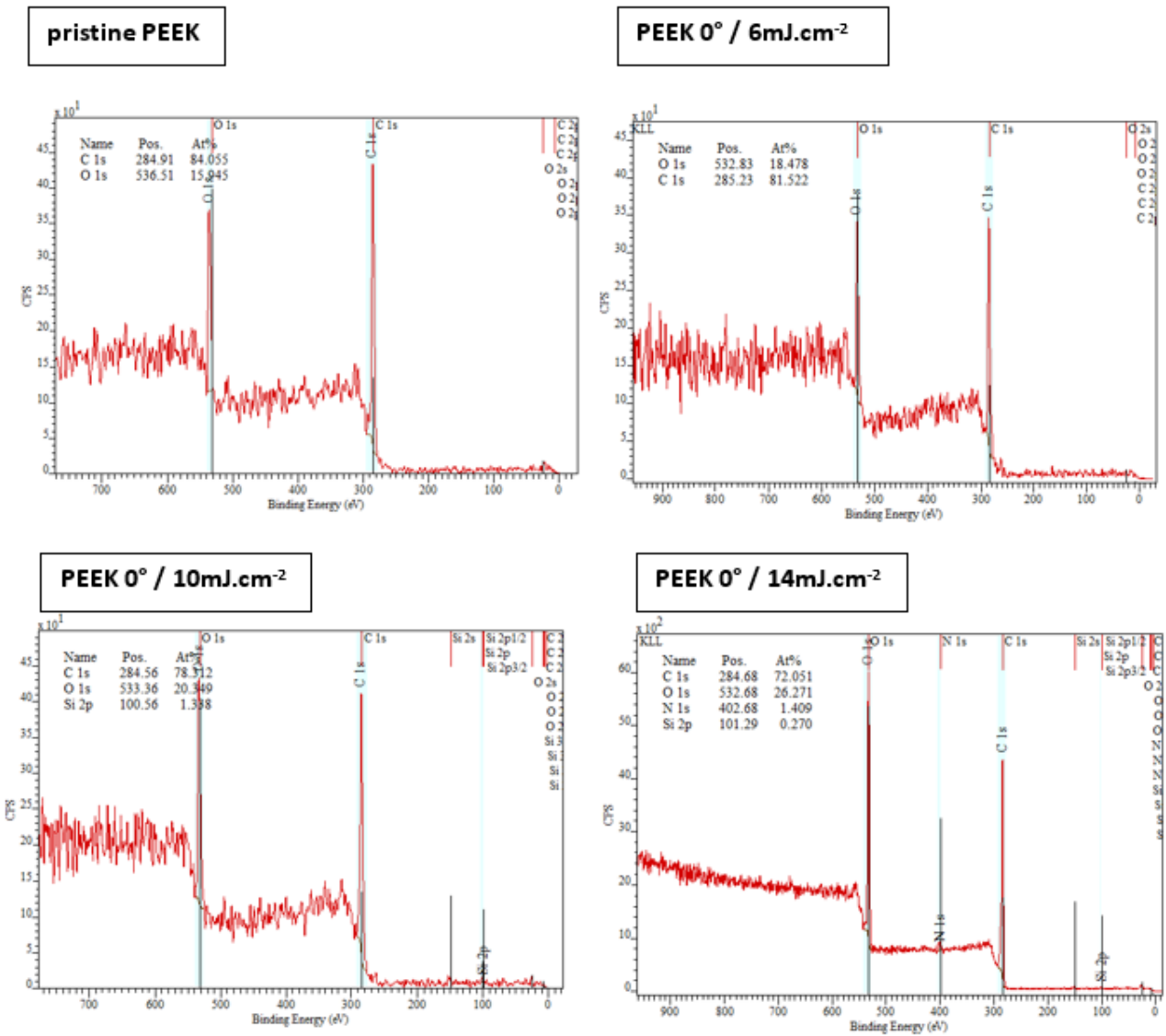

However, the increase of oxygen is much lower compared to PEN.

**Figure 8** XPS elemental spectra of the pristine PEEK surface and PEEK surface after laser modification with a laser fluence of (a) 6 mJ·cm$^{-2}$, (b) 10 mJ·cm$^{-2}$ and (c) 14 mJ·cm$^{-2}$.

**Table 2** Element concentration (at.%) determined by EDS analysis on PEEK and PEN polymers modified by laser fluence in the range of 4–16 mJ·cm$^{-2}$ (6000 pulses). The laser modification took place both single-phase at angles of 0° and 45° and two-phase (cross mod.). Measured area is 10 × 10 μm$^2$. The value highlighted in red (*) indicates the highest increase in atomic oxygen representation for the observed substrate with respect to the given type of modification.

| PEEK | 0° | | 45° | | cross mod. | |
|---|---|---|---|---|---|---|
| $E$ [mJ·cm$^{-2}$] | C [at.%] | O [at.%] | C [at.%] | O [at.%] | C [at.%] | O [at.%] |
| 0 | 87.9 | 12.1 | 87.9 | 12.1 | 87.9 | 12.1 |
| 4 | 87.6 | 12.4 | 84.9 | 15.1 | 86.8 | 13.2 |
| 6 | 85.5 | 14.5 | 86.1 | 13.9 | 87.3 | 12.7 |
| 8 | 87.0 | 13.0 | 84.2 | *15.8 | 87.5 | 12.5 |
| 10 | 85.0 | 15.0 | 84.4 | 15.6 | 85.6 | 14.4 |
| 12 | 88.4 | 11.6 | 87.8 | 12.2 | 85.3 | *14.7 |
| 14 | 83.3 | *16.7 | | | 86.3 | 13.7 |
| 16 | 86.3 | 13.7 | | | 87.0 | 13.0 |

| PEN | 0° | | 45° | | cross mod. | |
|---|---|---|---|---|---|---|
| $E$ [mJ·cm$^{-2}$] | C [at.%] | O [at.%] | C [at.%] | O [at.%] | C [at.%] | O [at.%] |
| 0 | 90.5 | 9.5 | 90.5 | 9.5 | 90.5 | 9.5 |
| 4 | 82.9 | 17.1 | 82.3 | 17.7 | 86.4 | 13.6 |
| 6 | 81.8 | 18.2 | 79.3 | 20.7 | 86.2 | 13.8 |
| 8 | 86.2 | 13.8 | 82.1 | 17.9 | 81.3 | *18.7 |
| 10 | 81.5 | *18.5 | 80.3 | 19.7 | 81.6 | 18.4 |
| 12 | 83.9 | 16.1 | 78.7 | *21.3 | 81.3 | *18.7 |
| 14 | 84.9 | 15.1 | | | 87.9 | 12.1 |
| 16 | 87.2 | 12.8 | | | 85.1 | 14.9 |

* highest value

The results of XPS analysis show the effect of different laser fluences on PEEK polymer surfaces after two-phase excimer exposure. Following the results of the EDS analysis, no significant difference in the surface oxygen concentration was detected using XPS, which was in the range of 18.9–22.3% for the selected samples. The wetting contact angle (θ) is one of the few directly measurable properties of the solid-liquid-gas three-phase interface. The equilibrium condition at such an interface is determined by Young's equation (3), which describes the dependence of the contact angle value on the size of the interphase energies at the solid-liquid interface. The dependence of contact angles on the energy of the laser flow (Figure 9) demonstrate that laser modification of the surfaces of polymer films significantly affects their wettability. This phenomenon can be explained by an increase in the number of oxygen (polar) groups on the surface of the polymers due to laser exposure.

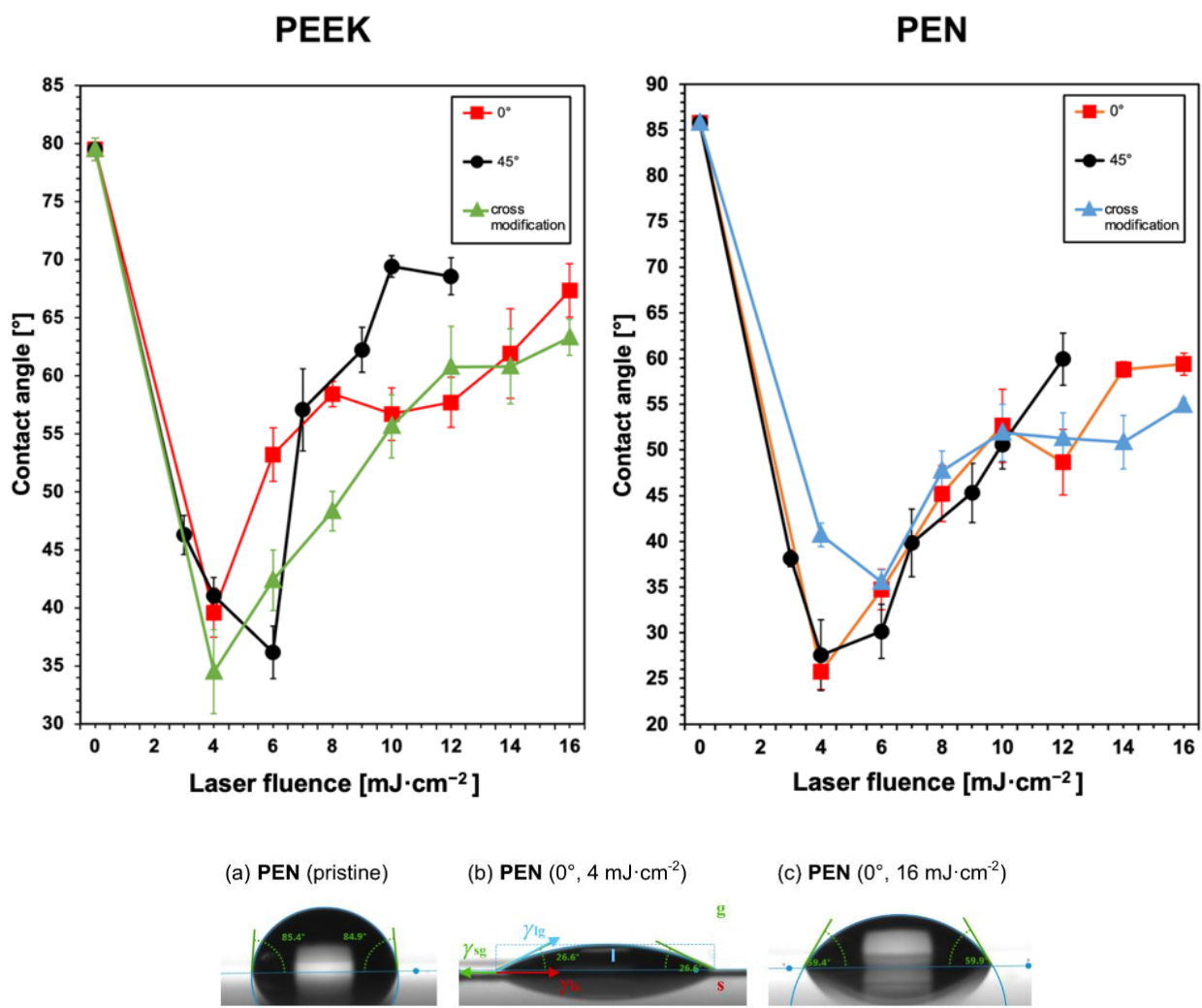

**Figure 9** Dependence of the contact angle of water on PEEK and PEN foils modified by laser fluence in the interval 4–16 mJ·cm$^{-2}$. The influence of single-phase exposure, at selected angles (0° and 45°), and two-phase (cross) exposure is observed. At the bottom the appearance of distilled water droplets on selected PEN surfaces. (a) Unmodified, (b) one-time modified surface at 0° with 4 mJ·cm$^{-2}$ laser fluence and (c) 16 mJ·cm$^{-2}$. The applied number of laser pulses was 6000.

It is evident, that the oxygen containing groups are induced immediately after the treatment, the applied laser fluence 4 mJ.cm-2 induced sharp decrease of contact angle for PEEK from ca 80° to 40°, for samples exposed by the cross modification the decrease was down to 35°, and even to 25° (PEN, perpendicularly exposed, 0°). These values may be considered as the lowest contact angle values. If we discuss the wettability, contact angle and surface chemistry, we see that there is not a clear dependence. More specifically, with further increasing oxygen concentration on the samples (see Figure 8 and Table 2) the contact angle remained the same, of it values increases. We would expect the wettability increase with increasing oxygen concentration, but we have to consider that there is a increase/change in oxygen concentration, due to mass transfer connected with LIPSS formation or globular pattern formation, the most upper layers, are "mixing" with the non-affected bottom layers. This decreased the surface radicals, carbonyl and carboxyl groups on the very surface, thus the contact angle increase for larger laser fluences.

The goniometry results further show that the wettability of the samples is affected by the angle of incidence of the laser beam, with the polymer samples that were irradiated at the angle of 45° showing greater variability in the values of the contact angle with water than the samples irradiated at the angle of 0°. As a rule, the lowest contact angle values were achieved by samples that were modified with a laser fluence of 4 mJ·cm$^{-2}$. The only exception was the PEEK foil sample (modified at an angle of 45°), which reached the lowest value at a laser fluence of 6 mJ·cm$^{-2}$ and a two-phase modified PEN foil sample (also 6 mJ·cm$^{-2}$). Gradually increasing the laser fluence led to an increase in the contact angle of all preparations and, thus, also to an increase in the hydrophobicity of their surfaces. The appearance of such surfaces is shown in Figure 9, bottom images. However, the increased values of the contact angles never exceeded the original value of the unmodified samples. A further increase in contact angle values could occur due to the so-called aging of the surface when the surface of modified polymer films is exposed to atmospheric conditions, which can cause changes in the chemical composition of the surface by reorienting the formed functional groups or radicals back into the volume of the material or by releasing oxygen from the surface of the sample.

Further, infrared spectroscopy (FTIR) was also used to study changes in the chemical composition of PEEK and PEN surfaces after laser exposure, the results of which confirmed the chemical identity of the samples. When discussing the changes in FTIR spectra, we have to consider laser irradiation volume and the potential interaction to the environment. Characteristic absorption bands were noted for the unmodified samples, such as those associated with C=O bonds and bands associated with vibrations of C=C aromatic rings. Figure 10 shows the difference spectra of (a) PEEK and (b) PEN substrates, which follow the changes caused by single-phase (0°, 45°) and cross-excimer modification. For all measured PEEK and PEN samples, the absorption bands with a wavenumber of ca. 1,725 cm$^{-1}$, which correspond to vibrations of the carbonyl (C=O) functional group, and bands with a wavenumber of ca. 750 cm$^{-1}$ of out-of-plane vibrations of C–H bonds were observed. Due to formation of LIPSS on the polymer surface, we have to consider the influence of surface modified layer with periodic nanopattern, and its contribution to the FTIR spectra of the sample as a whole. The oxygen containing groups induced by excimer treatment on the treated surface are both for PEN and PEEK polymer distributed not deeper that several hundreds of nanometers, more specifically are on the very top of the formed LIPSS. Therefore the contribution to the FTIR spectra depend on the altered thickness, and it could be supposed that the higher the LIPSS are, and the deeper affected layer by the excimer is, it will have more influence of the FTIR results.

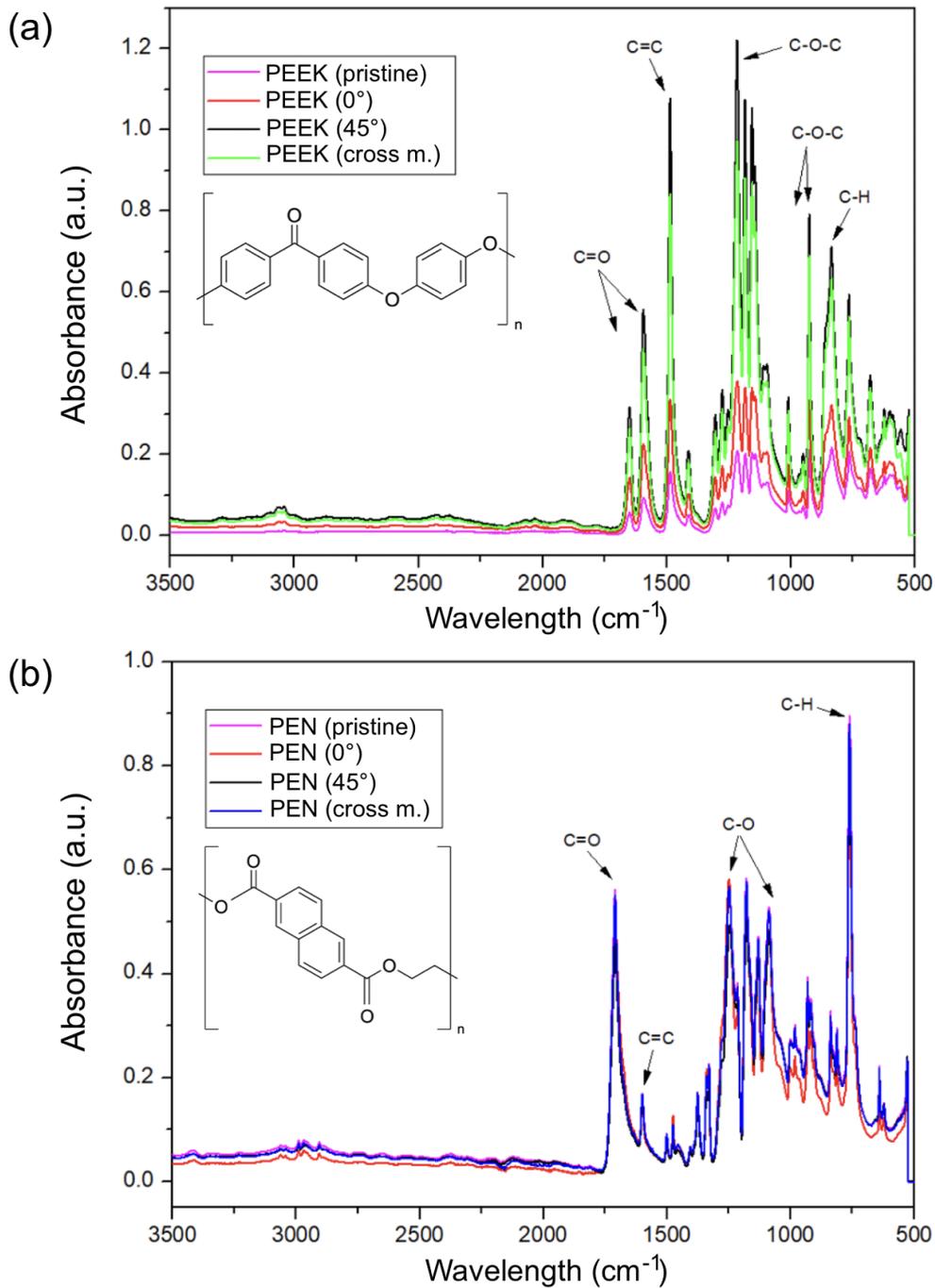

**Figure 10** FTIR spectra of unmodified and modified films (a) PEEK, (b) PEN by single-phase exposure at 0° and 45° and two-phase (cross) laser exposure (10 mJ·cm$^{-2}$). The applied number of laser pulses was 6000.

In the case of the PEN foil samples, it can be stated at first glance that there were no major changes in the maxima of the absorption bands. Also, only a small peak with a wavenumber value of 1,600 cm$^{-1}$ is observed due to the vibrations of the C=C double bonds in the aromatic rings of the functional naphthalene groups of the polymer chain. In contrast to the PEN difference spectra, a strong maximum of ca. 1,500 cm$^{-1}$ was captured on the PEEK substrates indicating the vibrations of the aromatic ring (C=C). The peaks captured around wavenumbers ranging from 1,275 to 1,000 cm$^{-1}$ most likely belong to the asymmetric vibrations of the ether (C–O–C) bonds occurring in the PEEK structure and the vibrations of the C–O bonds in the ester (–COO–) groups of the PEN polymer chain.

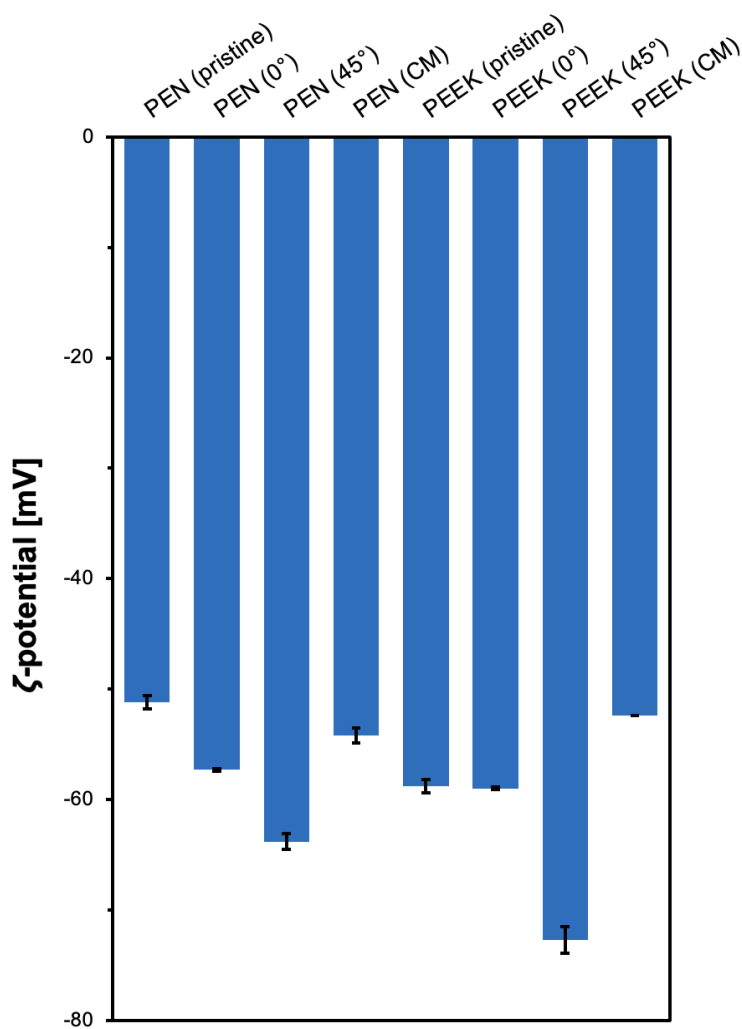

**Figure 11** Zeta-potential (ζ) values of unmodified (pristine) and modified foils (PEEK, PEN) by one- and two-phase (cross) laser exposure (CM) with a laser flux of 10 mJ·cm$^{-2}$ at an angle of 0° and 45°. The applied number of laser pulses was 6000.

In order to characterize and observe changes in the electrical properties of polymer surfaces, the electrokinetic potential ζ was also measured. Zeta potential indicates the presence of functional groups on the material surface. For instance, an increase in the number of oxygen-containing functional groups on the polymer surface can lead to an increase in the zeta potential towards positive values, while a decrease in surface hydrophilicity can lead to a zeta potential decrease. The measurement of the ζ-potential is particularly sensitive to the pH, ionic strength of the liquid phase, material porosity, conductivity, and surface roughness, among others. In this study, the measurement took place only at pH~6, and samples modified with a laser fluence of 10 mJ·cm$^{-2}$ were selected. Interestingly, the data in the graph in Figure 11 do not correlate with the results of the EDS analysis, in which it was found that laser modification at the angle of 45° led to the highest values of atomic representation of oxygen on the PEEK and PEN surface.

A higher atomic representation of oxygen may be related to the more abundant presence of oxygen groups, such as –OH, –C=O, and –COOH, on the material surface, which may promote an increase in the zeta potential value to positive values. This is because the oxygen functional groups can attract oppositely charged ions from the solution, creating an electrical double layer of charged ions on the surface. The ions of the outer mobile phase of the electrolytic double layer thus have the opposite sign than the sign ζ-potential. In our case, however, on the contrary, a reduction of the zeta potential to negative values was observed, the values of which were the most negative for the samples modified by the laser flux at the angle of 45° (for PEEK and PEN -72.7 and -63.8 mV, respectively). The main reason for the change of zeta potential to more negative values will most likely be the high roughness of these surfaces. Due to the significant roughness, the geometry of the surface is significantly affected, resulting in the formation of an inhomogeneous electrical double layer. When the electrolyte flows around the sample, the effect of uneven surface geometry can lead to an increase in the concentration of charge carriers inside the return flow regions. This can result in a disruption of the surface charge leading to an overall decrease in the zeta potential.

*3.3 Study of the cellular response to the surfaces of selected substrates*

As a next step, we aimed to evaluate the biocompatibility of the laser-treated PEEK and PEN polymer samples using C2C12 cells. For this measurement, we selected samples that were treated with a laser fluence of 10 mJ·cm$^{-2}$ (6,000 pulses) single-phase at the angle of 0° and 45°, and two-phase (KM). As the control substrate standardly used for mammalian cell cultivation, tissue culture polystyrene (TCPS) was chosen. To determine the cell number and homogeneity of cell distribution, the cell cytoskeleton was labeled with phalloidin conjugated to Atto488 and nuclei with DAPI, see fluorescence microscopy images in Figure 12. Regrettably, high autofluorescence of PEEK, as already reported by us [63] and PEN samples slightly affected the image quality. Possibly, in the future, this strong autofluorescence could be reduced by using UV irradiation of the samples before cell culture. From Figure 13 it is obvious, that the cell alignment is as anticipated, the cell are aligned along the pattern, and the increasing pattern width has the influence on the cell placement.

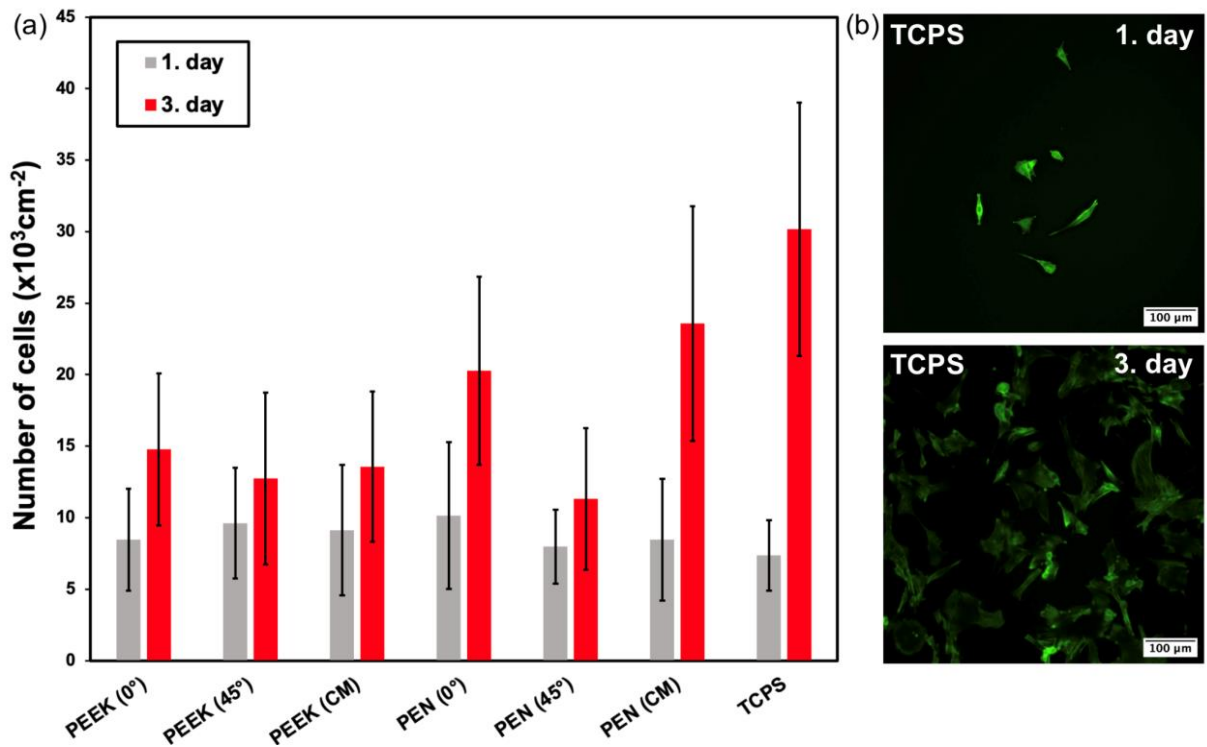

**Figure 12** (a) Graph showing the number of adhered (day 1) and proliferated (day 3) C2C12 cells cultured on laser-treated PEEK and PEN samples at a fluence of 10 mJ·cm$^{-2}$ compared to TCPS. The effect of angle modification (0° and 45°) and cross modification (CM) was also tested. (b) Fluorescence microscope images of cells on TCPS substrate at day 1 and day 3.

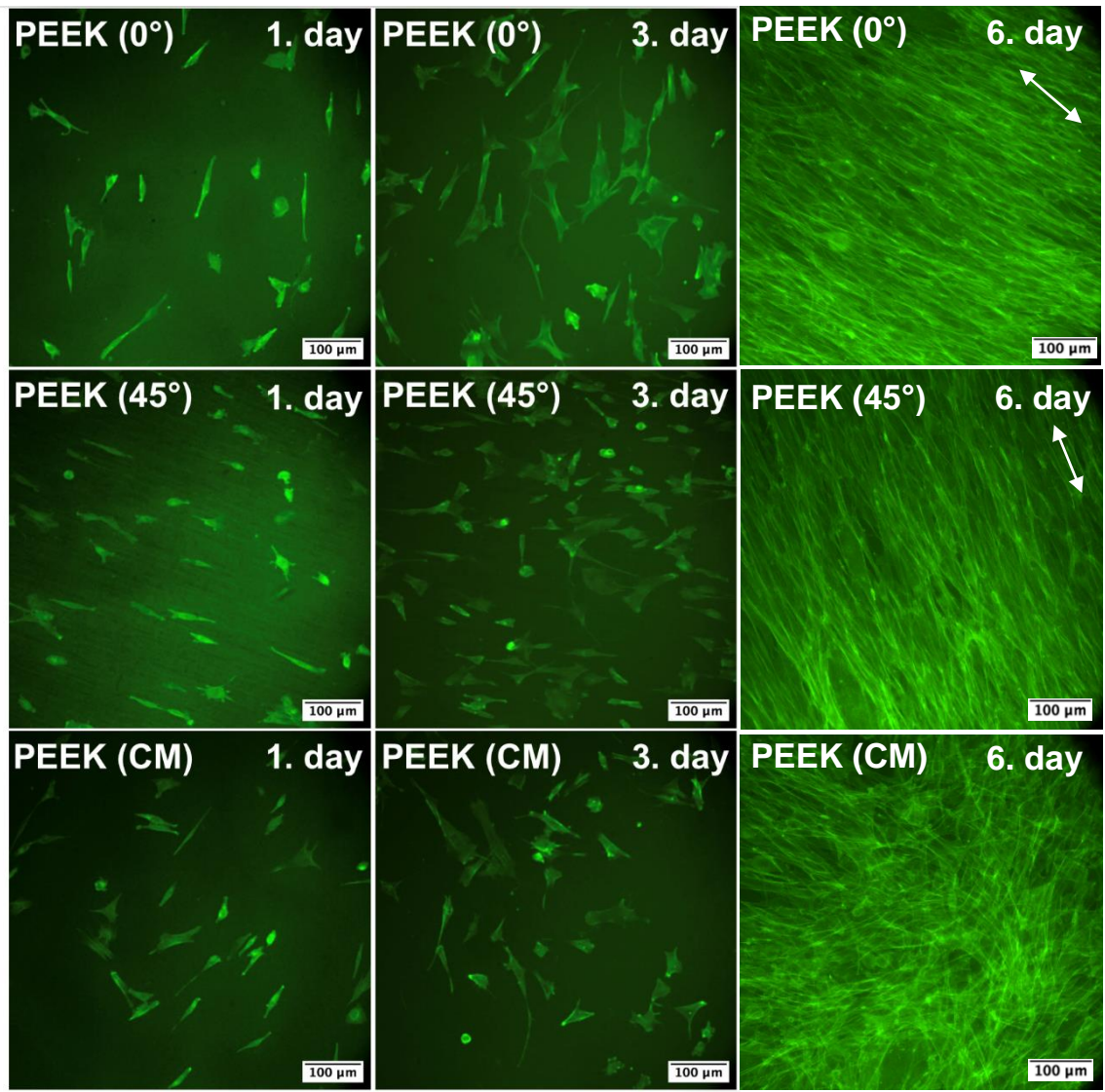

**Figure 13** Images of adhered (day 1) and proliferated (day 3, day 6) C2C12 cells cultured on laser-modified PEEK (0°, 45°, CM) substrates. The arrow indicates the ripple alignment.

We see that the cell alignment follows the pattern shape/direction induced on the PEEK by the excimer laser. The cell alignment is better visible on samples analyzed after 6 days from seeding, the cells are aligned along the pattern diretion. This alignment is visible both for samples treated perpendicularly (0°), and under larger angle of incidence (45°), as it is evident from the 2nd line of Figure 13. If cross modification is applied, the cell formation is no longer aligned, but rather "cross pattern of cell placement" was observed. It is obvious, that the cytocompatibility of the PEEK laser treated surface is excellent.

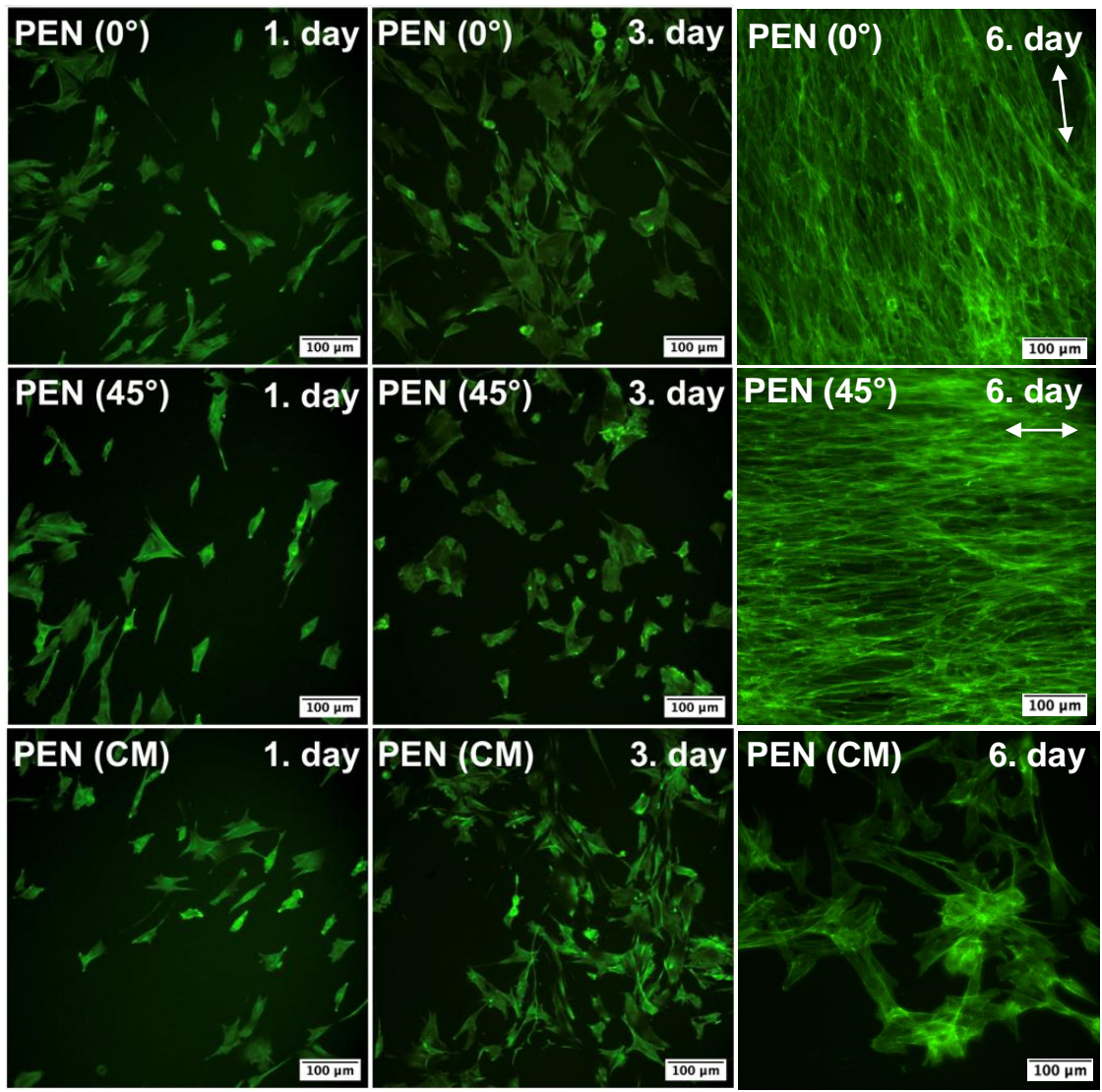

**Figure 14** Images of adhered (day 1) and proliferated (day 3, day 6) C2C12 cells cultured on laser-modified PEN (0°, 45°, CM) substrates. The arrow indicates the ripple alignment.

We see that the cell alignment follows the pattern shape/direction induced on the PEN by the excimer laser. The cell alignment is better visible on samples analyzed after 6 days from seeding, the cells are aligned along the pattern diretion. This alignment is visible both for samples treated perpendicularly (0°), and under larger angle of incidence (45°), as it is evident from the 2nd line of Figure 14, the cell are „better" aligned on the sample with larger period, which in induced by larger angle of incidence (PEN 45°). From Figure 14 it is obvious, that the cells are aligned along the pattern, and the increasing pattern width has the influence on the cell placement. If cross modification is applied, the cell formation is no longer aligned, but rather "cross pattern of cell placement" was observed. But, compared to PEEK substrate, the cell number and area cover is significantly decreased.

It is apparent that 1 day after seeding, the cells were homogeneously distributed on the surfaces of all tested PEEK and PEN nanostructured samples. On average, cells adhered to the samples at initial cell population densities of 9,077 ± 3,997 and 8,860 ± 3,988 cells per cm$^2$ for PEEK and PEN substrates, respectively. The lowest number of adhered cells was found on the TCPS, however, this changed on day 3 when a significantly higher number of cells (30,000 ± 9,000 cells·cm$^{-2}$) were detected on TCPS compared to the rest of the evaluated samples. At three days of cultivation, the number of cells on all types of samples increased, however, the cells on the nanostructured PEEK and PEN surfaces were less homogeneously distributed and formed clusters. The inhomogeneous cell coverage on these substrates was probably caused by the relatively high roughness of their surfaces compared to commercial TCPS, the mean roughness of which corresponds to ca. 5.4 nm [64]. The cell population density most closely approximated the TCPS results on PEN (KM) samples that were modified by biphasic exposure.

## 4. Conclusions

Here, we have demonstrated that the use of a KrF excimer laser of 248 nm is an effective method for creating periodic pattern on the PEEK and PEN surfaces. Polymer foils were exposed to laser fluences at 4–16 mJ·cm$^{-2}$ (number of laser pulses 6,000) to minimize the negative effects of ablation and thermal damage to the material. Excimer exposure of both PEEK and PEN induced nanometer-scale periodic waves with different periodicity and height. In addition to wave-shaped nanostructures, dotted and pod-like patterns were also prepared. The optimal laser parameters for the reconstruction of LIPSS nanostructures were determined for higher laser fluence above 6 mJ·cm$^{-2}$. The maximum energy of the laser beam for the formation of LIPSS was found at a fluence of 12–14 mJ·cm$^{-2}$. Above this value, i.e. an energy of 16 mJ·cm$^{-2}$, a gradual distortion of the periodic structures already occurred.

From the point of view of the chemical composition of the surface of the prepared samples, it was proven that changes in the atomic representation of the C and O elements occur due to the laser exposure; in almost all cases an increase in the O percentage was detected. The lowest contact angle values were achieved by samples that were modified with a laser fluence of 4 mJ·cm$^{-2}$. The average values of these contact angles were 38.4° for PEEK surfaces and 31.3° for PEN across all types of laser modification. The increase of the laser fluence led to an increase in the hydrophobicity of the treated surfaces, the effect was accompanied with an increase in the mean surface roughness value. Higher surface roughness also led to a decrease in the zeta potential value to more negative values. Further, infrared spectroscopy (FTIR) confirmed the chemical identity of the samples, with characteristic absorption bands such as those associated with C=O bonds and C=C aromatic rings.

Cytocompatibility using C2C12 cells was demonstrated using cell counting and morphology examination by fluorescence microscopy. The impact of globular/dot structures on the cell growth in comparison to pristine or linear LIPSS-patterned surfaces was determined. The linear pattern (LIPSS) induced the cell alignment along the pattern direction, while its cytocompatibility has been enhanced. The dot/globular pattern had no such effect such as alignment, but the cell number was similar (PEEK) or higher (PEN) compared to polymer samples with LIPSS only. The PEN cross modification exhibited the highest cell numbers compared to tissue polystyrene.

**Data availability**

The data presented in this study are available at https://doi.org/10.5281/zenodo.10444586.

**Acknowledgments**

This work was supported from the grant Czech Science Foundation under project 21-05506S and Project OP JAK_Mebiosys, No CZ.02.01.01/00/22_008/0004634 of the Ministry of Education, Youth and Sports, which is co-funded by the European Union.